\documentclass[prl,twocolumn,showpacs,floatfix]{revtex4} 

\usepackage{graphicx}
\begin{document}

\title{Kinetic pinning and biological antifreezes}
\author{Leonard  M. Sander and Alexei V. Tkachenko}
\affiliation{Michigan Center for Theoretical Physics and Department of Physics,
	University of Michigan, Ann Arbor, Michigan, 48109-1120}
date{\today}

\begin{abstract}
Biological antifreezes protect cold-water organisms from freezing. An example are the antifreeze proteins (AFP's) that  attach to the surface of ice crystals and arrest growth. The mechanism for growth arrest  has not been heretofore understood in a quantitative way. We present a complete theory based on a kinetic model. We use the `stones on a pillow'  picture  \cite{Knight01b, Knight91}. Our theory  of the suppression of the freezing point as a function of the concentration of the AFP is quantitatively accurate.
It gives a correct description of the  dependence of the freezing point suppression on the geometry of the protein, and might lead to advances in design of synthetic AFPÕs.
\end{abstract}

\maketitle 
 In polar regions, many fish, insects and plants flourish at temperatures
well below the freezing point of their bodily fluids \cite{Duman01,Fletcher01}.
Often, particularly in fish \cite{Devries83},  this is a non-colligative effect, 
namely that  ice crystal growth
within the organisms is arrested by a class of plasma proteins called
anti-freeze proteins (AFP). They are usually
peptides or glycopeptides. These molecules attach irreversibly to ice
surfaces \cite{Devries83} and lead to arrest of crystal growth\cite{Knight01a,Knight01b}
until the water containing the AFP's is supercooled by as much as 2 degrees
C. The mechanism for the suppression of the freezing point is not
understood. In this paper we present a growth model allows us to understand
AFP's in considerable detail, and, in particular to
calculate the dependence of the undercooling on AFP concentration. 
\cite{Chao97,Deluca98}.

Growth arrest by AFP's occurs because
the protein adsorbs on the surface of the growing ice crystal,
is not incorporated in it and suppresses growth at that site. 
One version
of this notion has been called the `stones on a pillow model' 
\cite{Knight01b,Knight91}. It assumes a thermally rough crystal,  appropriate
for most of the surface of ice near its freezing point. (The basal plane of ice is facetted,
but AFP's adsorb mostly on the prism planes \cite{Knight01a}.) The AFP's
are  obstacles to growth so that the
crystal surface bulges between the attachment sites, and
the freezing point is depressed by the well-known Gibbs-Thompson effect,
namely that a curved surface has a lower freezing point than a flat one; see
Figure (1a).
This is the physical picture we will adopt. The surface of the
crystal which is not under the AFP must have constant  mean curvature, $\kappa$.
The Gibbs-Thompson condition \cite{Langer80} is:
$\kappa = -\delta_T/l_o $
where $\delta_T=(T_m-T)/T_m$ (the undercooling) and $T_m$ is the
equilibrium melting temperature \cite{Langer80}: 
Also, $l_o=\gamma/\Lambda$ is a characteristic length, $\gamma$ the
interfacial tension, and $\Lambda$ is the latent heat of fusion per unit
volume. For ice $l_o \approx 1 \mathrm{\mathring{A}}$.

Consider a surface, $h(x,y)$ which has stopped growing. We model
the AFP's as spheres of radius $b\sim 10\mathrm{\mathring{A}}$.
(We treat more complicated molecular geometries, below.) Assume
that the particles are half buried in the ice, and fix the ice surface at
the equatorial plane of the sphere. The angle of the ice with the equatorial
plane can take on any value. We 
use the small slope approximation, $|\partial h/\partial x|<<1,|\partial h/\partial y|<<1$, so that the
 curvature of the interface is given by  $\kappa\approx \Delta h$. Then 
 the Gibbs-Thompson condition becomes the Poisson
equation: 
\begin{equation}
\Delta h=-\delta _{T}/l_{o}\equiv -4\pi \rho .  \label{Poisson}
\end{equation}
This equation is familiar in electrostatics.  $h$ plays the role of the potential, and  the effective (positive) charge density, $\rho $ is defined by Eq. (\ref{Poisson}).

For simplicity, consider a periodic square array of AFP's. The
boundary condition on $h$ at the edge of the unit cell is that the
normal derivative vanishes, and at the edge of the AFP, $h=0$. This
 problem is easily solved numerically and a
contour plot  is shown in Figure (1b). In
order for a solution to the equation to exist, the AFP must act as
a negative charge so that the system is neutral. That is, $\rho =-nq$, where 
$n$ is the density of AFP's on the surface. By integrating 
Eq. (\ref{Poisson}) over the surface we find: 
\begin{equation}
\left\langle \left( \mathbf{r-r}_{i}\right) \cdot \nabla h\right\rangle
_{\left\vert \mathbf{r-r}_{i}\right\vert }=\delta _{T}/(2\pi l_{o}n)
\label{undercool}
\end{equation}
where $\left\langle \cdot \right\rangle $ is the average around the
edge of the AFP. The effective charge
on the AFP is: 
\begin{equation}
q=-(1/2)\left\langle \left( \mathbf{r-r}_{i}\right) \cdot \nabla
h\right\rangle _{\left\vert \mathbf{r-r}_{i}\right\vert }  
\label{charge}
\end{equation}

Note from Eq. (\ref{undercool}) that  the slope at
the edge of the molecule increases with undercooling. It is reasonable to
assume that if the slope exceeds some critical value, $\chi$,  the
antifreeze molecule will be engulfed.
Here $\chi$
is set by the physical chemistry of the AFP and the
interface. We must have  $\chi \approx 1$.
The maximum undercooling,  $\delta _{T}^*$, is given by Eq. (\ref{undercool})
with $<\partial h/\partial r>=\chi $. 
For $b\sim 10\mathrm{\mathring{A}}$ , and $\delta _{T}^*$ of order
1 degree we need $n=n^{\ast }=\delta _{T}^*/(2\pi \chi l_{o}b)\approx 6%
\mathrm{x}10^{11}\mathrm{cm}^{-2}$, or a distance between AFP's of order 100$%
\mathrm{\mathring{A}}$. From Eq. (\ref{charge}) we find $-2q<\chi b < 0$.

\begin{figure*}[tbp]
{\parbox{3.5in}{ \includegraphics*[width=3.5in] {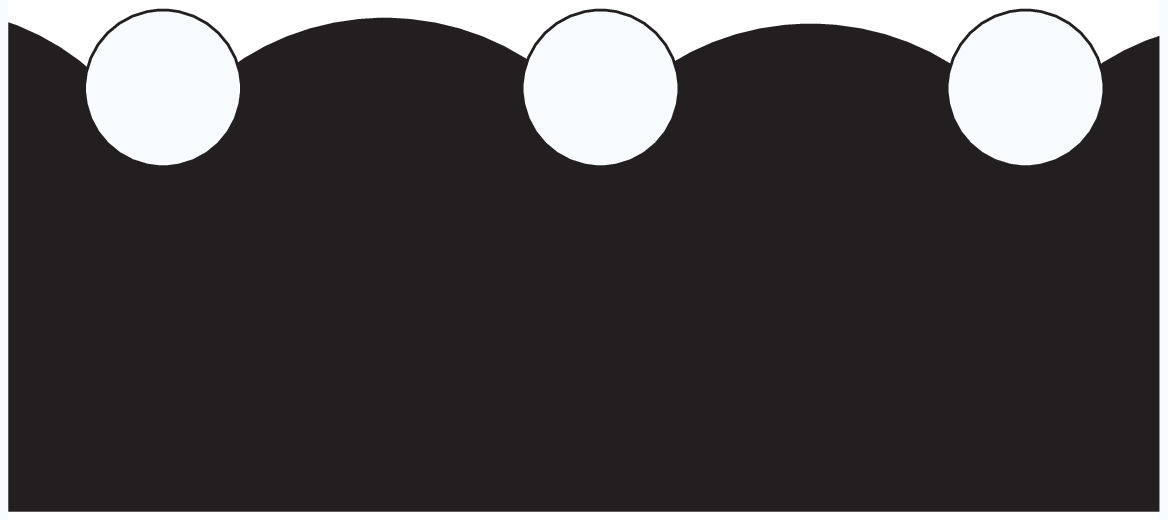}}}
\parbox{3in}{\includegraphics[width=3in]{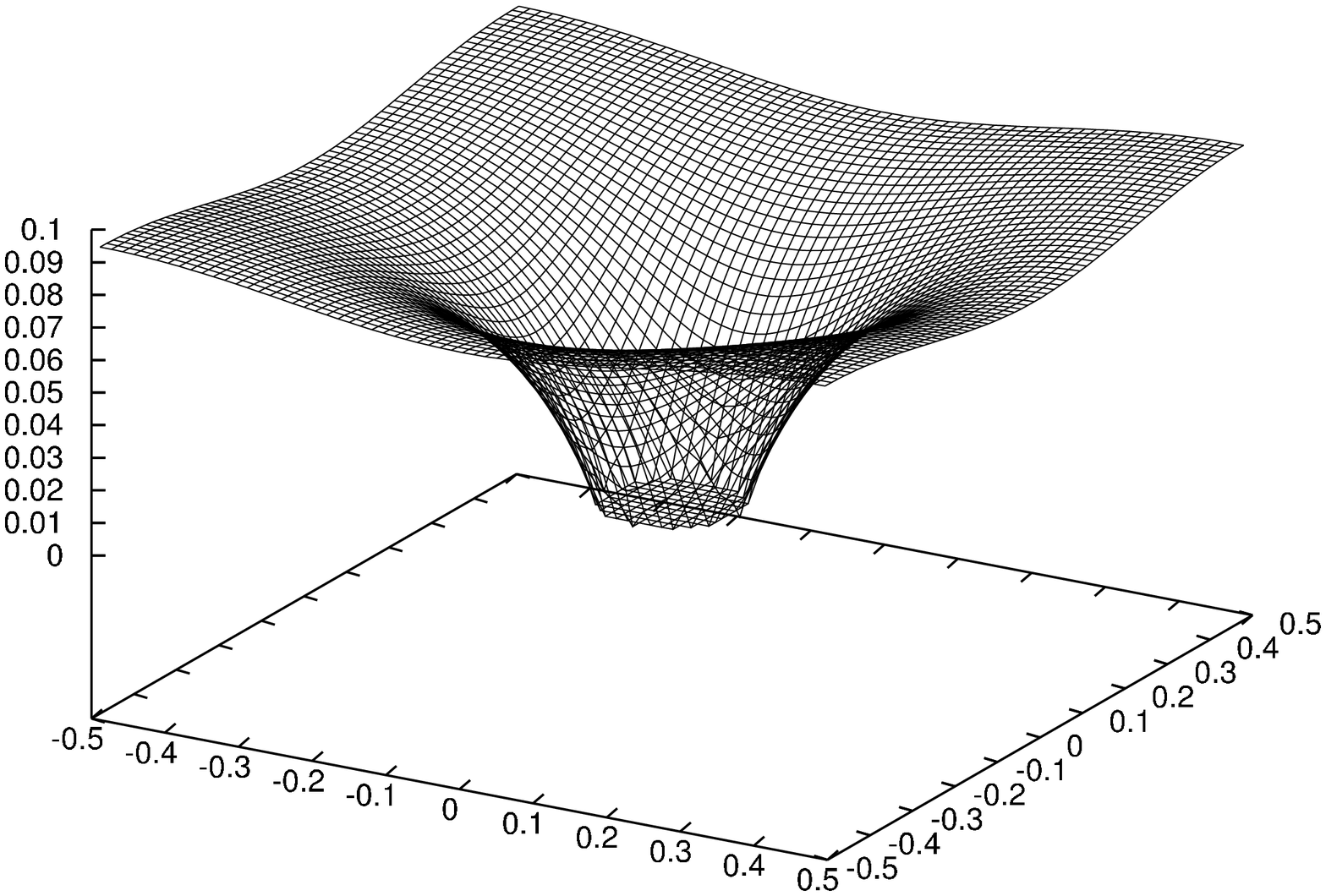}}
\caption{ a).The `stones on a pillow' model. AFP molecules, white circles,
attach to a surface. b). Visualization of a surface of constant mean
curvature. The size of the unit cell has been set to unity.}
\label{sketch}
\end{figure*}

Since the adsorption is thought to be irreversible, the
relationship between $n$, the density of AFP on the surface,  
and the concentration in solution  must depend on the kinetics of the growth
of the crystal, which we now model.
In the simplest picture, the growth speed of the crystal boundary  
is proportional to the variation of the overall free energy
of the system with respect to the normal displacement. 
In the small slope limit we take this to be $\delta h$. Then:
\begin{equation}
\dot{h}=-\Gamma  \frac{\delta }{\delta h}\int \left[ \gamma [1+|\nabla h|^2]^{1/2}+\Lambda
\delta_T h \right] d^{2}\mathbf{r.}  
\label{variation}
\end{equation}
The first term of the integrand is the surface energy, and the second
is the difference of the bulk free energies of solid and liquid
phases close to melting point. $\Gamma$ is a kinetic coefficient related to the rate of
attachment of water molecules to the ice surface.

From the calculus of variations \cite{Goldstein} this equation is equivalent to:
\begin{equation}
\dot{h}= v_{o}\left[ l_{0}\Delta h+\delta _{T}\right] .  
\label{diff}
\end{equation}
Here $v_{o}=\Gamma \Lambda $. This is a diffusion equation for $h$.
The AFP's are pinning centers for the surface, whose effect may be expressed
as a set of boundary conditions: 
\begin{equation}
\left. h\right\vert _{\left\vert \mathbf{r-r}_{i}\right\vert =b_{0}}=h_{i}.
\label{hi}
\end{equation}%
Here $\mathbf{r}_{i}$ is the position of $i$-th AFP and $h_{i}$ is the
height of its equatorial plane.

Suppose that the AFP's in the water adsorb at random on the
ice surface with rate, $k_{+}$, per unit area. The value of $h_{i}$ is the
local position of the interface at the moment when the $i$-th AFP is
adsorbed. The AFP's effectively
disappear when they get buried under the ice surface. 

It will be useful to split $h\left( \mathbf{r}\right) $ into two contributions: 
$h\left( \mathbf{r}\right) =H +\phi \left( \mathbf{r} \right)$,
where $H $ is the average height, and $\phi \left( \mathbf{r}\right) $
measures the small-scale variations. 
Suppose the AFP's are randomly distributed with density, $n$. We 
choose $\phi $ to develop in time as:
\begin{equation}
\dot \phi = v_o l_o( \Delta \phi + 4\pi\rho)
\label{phidot}
\end{equation}
where $\rho= n<q>$ is chosen to make the surface neutral, and $<q>$ is
the average charge. It is not hard to show that 
the time derivative in Eq. (\ref{phidot}) for $\phi$ is small (the quasi-static limit).  Thus:
\begin{equation}
\Delta \phi =-4\pi \left[ \rho +\sum\limits_{i}q_{i}\delta \left( \mathbf{r-r%
}_{i}\right) \right] .  \label{Poiss}
\end{equation}%
The point charges account
for the boundary conditions, Eq. (\ref{hi}). 
The values of $q_{i}$ are not fixed: as the interface moves,
the charges vary from $0$ at the moment of creation to
 $-\chi b/2$, at which point the AFP is engulfed by ice. 

 Near an
AFP, ($b<\left\vert \mathbf{r-r}_{i}\right\vert \ll 1/\sqrt{n}$), $\phi $ is
dominated by the contribution of a single charge \cite{Jackson}: 
\begin{equation}
\phi \left( \mathbf{r}\right) \simeq -q_{i}\log \pi \left( \mathbf{r-r}%
_{i}\right) ^{2}n  \label{log}
\end{equation}
Therefore: 
\begin{equation}
q_{i}\simeq \frac{H-h_{i}}{\log \left( \pi nb^{2}\right) }<0  
\label{charges}
\end{equation}
Since $|q|$ lies between 0 and $\chi b/2$, the `hole' in the interface
associated with the i$^{th}$ AFP is of order $|q_{i}|\sim \chi b/4$.
Subtracting Eq, (\ref{phidot}) from   Eq. (\ref{diff}), we find: 
\begin{equation}
\dot{H}=v_{o}\left[\delta _{T}-4\pi l_{0}\rho\right] .  \label{Hr}
\end{equation}
Suppose the interface moves with constant speed $V$. Since each 
$h_{i}$ is fixed and $H$ changes uniformly with time, Eq. (\ref{charges})
implies that the magnitudes of individual point charges $q_{i}$ are
uniformly distributed between $0$ and $-\chi b/2$, i.e. $\left\langle
q\right\rangle =-\chi b/4$ and $\rho =nb/4$. This implies that 
$V=v_{o}(\delta _{T}-\pi \chi l_{o}bn)$.

The evolution of  $n$ can be calculated by noting that its rate of increase is $k_{+}$, and that 
its rate of decrease is the rate that  AFP's are engulfed. 
 Eq. (\ref{charges}) implies
 $\dot{q_{i}}=V/\log (\pi nb^{2})$. Since the $q_{i}$ are uniformly distributed we must have
 $dn/n=-\dot{q_{i}}dt/(\chi b/2)$. This gives: 
\begin{equation}
\dot{n}=k_{+}-\frac{2nV}{\chi b\log \left( 1/\pi nb^{2}\right) }.
\label{ndot}
\end{equation}

In the steady state we put $\dot{n}=0$ in Eq. (\ref{ndot}), and use the expression for $V$.
Thus,
\begin{equation}
\frac{k_{+}}{v_{o}\delta _{T}}=\frac{n\left[ 2-n/n^{\ast }\right] }{\chi
b\log \left( 1/\pi nb^{2}\right) },
\end{equation}%
where $n^{\ast }=\delta _{T}\left( 2\pi \chi bl_{0}\right) ^{-1}$. 
The right hand side of this equation has a 
maximum  for $n\approx n^{\ast }$. Thus,
 \emph{there is no constant speed solution} unless the right hand side is small
enough, i.e. for small enough $k_{+}$ or large enough $\delta_T$. The
threshold $\delta _{T}^*$ obeys: 
\begin{equation}
 (\delta _{T}^*)^2/\log(2\chi l_{0}/b\delta _{T}^*)
 \simeq  
 (\chi b)^2 2\pi l_{0}k_{+}/v_{o}
 \label{res1}
\end{equation}

\begin{figure*}
{\includegraphics[width=5in]{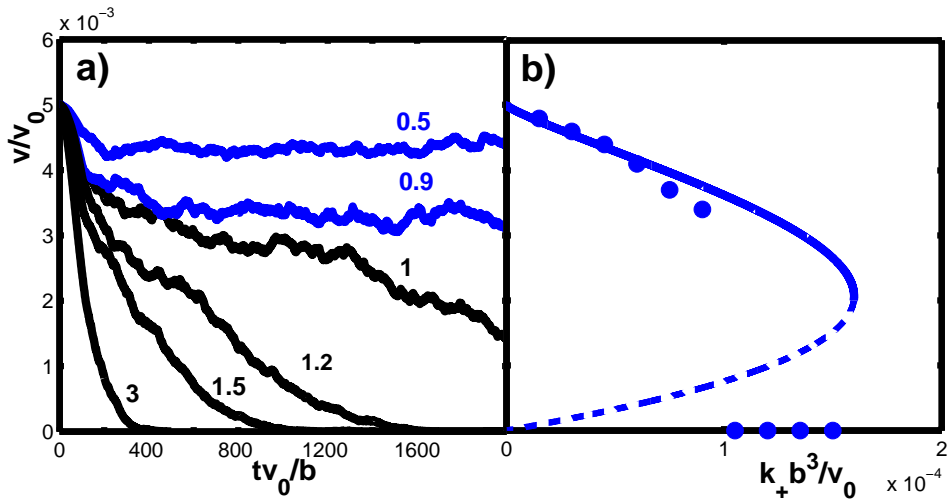} }
\caption{ (a) Simulation results for the speed
of the interface as a function of time, for various values of  control
parameter $k_{+}b^{3}/v_{0}$, ranging from $0.5\times 10^{-4}$ to $3\times
10^{-4}$.  $\delta_T=0.005, \chi=1/(2\pi)$
(b) Speed in the steady state vs. reduced adsorption rate:
comparison of  simulation (points) with  analytical result (line)
given by Eq. \protect\ref{res1}. }
\label{icefig}
\end{figure*}

To test our approximations, notably the quasi-static limit for $\phi$, we have performed a numerical simulation of  Eqs. (\ref{diff}), (\ref{hi}),
coupled to the random adsorption of AFP. Results
 on a $150\times 150$ square lattice, with each cell
representing a single AFP, are shown in
Figure \ref{icefig}. These results
support our analysis. In fact,
the transition from the steady growth to the arrested interface regime
occurs at a somewhat lower adsorption rate $k_{+}$ than predicted.
Since we have not performed a complete stability analysis
of the steady-growth solution, Eq. (\ref{res1}) gives only an upper bound
for the critical value of $k_{+}$. 
However, the discrepancy is rather small.

Beyond the transition point, growth stops. The resulting
 static interface must obey Eq. (\ref{diff}) with $\dot{h}=0$. Once
again, we recover the Poisson equation, Eq. (\ref{Poisson}).  After arrest, as 
Eq. (\ref{ndot}) shows, $n$ increases as irreversible adsorption continues until
limited by some aspect of surface chemistry that we have not considered in
our model.

Real AFP's are often anisotropic. In order to account for this, we assume that
the region blocked is  elliptical, with semi-major and semi-minor
axes $a$ and $b$, respectively.  The potential  $h\left(x,y\right) $ 
 can be found by  conformal mapping \cite{Churchill}. Using this method, it is easy to show that
 Eq. (\ref{charges}) is replaced by:
$H-h_{i}\simeq -q_{i}/\log \left( \pi n\left( a+b\right) ^{2}/4\right) $. 
The slope of $h$ will reach its critical
value near $x=\pm a$, and  its maximum  will be
given by the radius of curvature of the ellipse, of order $b$. Thus $q_{\max }=\chi b/2$. 
With these changes, the left-hand side of Eq. (\ref{res1}) becomes
\begin{equation}
(\delta _{T}^*)^2/\log(8\chi l_{0}b/(a+b)^2\delta _{T}^*)
\label{aniso}
\end{equation}

Eq. (\ref{res1}) involves  $k_+/v_{o} $ 
The rate, $k_+$ is certainly proportional  to the concentration, $C$, of 
AFP in solution: $k_+ = v_AC$ where $v_A$ has the units of velocity.
Therefore, $v_A/v_{o}$ is a  parameter  which we need to
estimate. 

In any experiment,  soon after the adsorption process starts water near
the ice surface is depleted of AFP. The thickness of the depletion layer
depends on the diffusion coefficient $D$, as $\sqrt{Dt}$.
That is, $k_{+}$ becomes time--dependent, and thus
$v_{A}\simeq \sqrt{D/t}$. The same argument applies to the rate of
crystallization itself: our estimate for $v_{o}$ fails as soon as the
process is limited by thermal diffusion through a diffusive
layer. Thus $v_{o}\simeq \sqrt{\kappa _{T}T/\Lambda t}$, where $\kappa
_{T}$ is thermal conductivity of water. From the Einstein
formula  $D=k_{B}T/6\pi \eta b$:
\begin{equation}
v_{A}/v_{o}=\sqrt{\lambda/b}\sim 10^{-2}  \label{ratio}
\end{equation}
where $\lambda =k_{B}\Lambda /6\pi \eta \kappa _{T}=1.2\times 10^{-10}$ cm.
By combining Eqs. (\ref{res1}) and (\ref{ratio}) we can write:
\begin{equation}
\delta _{T}^*\left[ \log (2\chi l_{0}/b \delta _{T}^*)\right]
^{-1/2}
\simeq
 \chi \sqrt{2\pi b^{3/2}l_{0}\lambda ^{1/2}C},  \label{deltaT}
\end{equation}

For anisotropic AFP's we must  use  Eq. (\ref{aniso})
and also note  that the longer axis,  $a$,  sets the
hydrodynamic radius of the ellipsoidal particle. Its
diffusion coefficient may be approximated by $D\approx k_{B}T\left(\beta+1\right) /6\pi \eta
a$, where $\beta=\log((a+b)/2b)$ (typically, $\beta < 2$). Thus 

\begin{eqnarray}
\delta_{T}^*[ \log \left( 2\chi l_{0}/b\delta _{T}^*\right)-2\beta ]^{-1/2}   \nonumber \\
\simeq \chi b\left( \frac{\lambda }{a}\right) ^{1/4}\sqrt{2\pi \left(
\beta +1\right) l_{0}C},  
\label{DeltaT1}
\end{eqnarray}

%\begin{widetext}
%\end{widetext}

\begin{figure}
{\includegraphics[width={3.5in}]{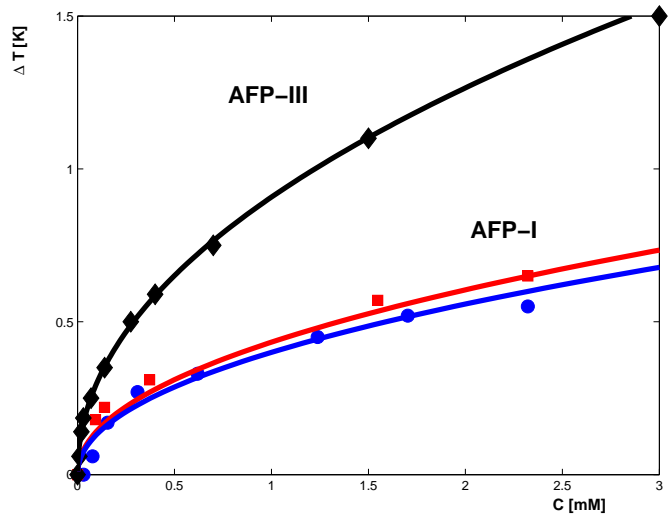}}
\caption{Comparison of theory to  experiment
 for  maximum undercooling \ of water in the presence of \  
AFP--Type III (diamonds) and two modifications of AFP--Type I \ (circles,
squares).} \label{afp}
\end{figure}

Using these formulas, we have been able to fit the experimental data
on natural AFP's of two different classes: AFP--type I, and
AFP--type III \cite{Chao97,Deluca98}. They have rather different architectures: AFP I is an 
$\alpha $-helical rod-like molecule which we model as a thin cylinder
 ($b=3.5\mathrm{\mathring{A}}$ , $a=25\mathrm{\mathring{A}}$), while AFP III has a more
complicated globular structure, which we approximate as a sphere
 ($b=a=8\mathrm{\mathring{A}}$). Note that these lengths are derived
from the known structures of the molecules,
and therefore the only free parameter of our theory is $\chi$. 
The theoretical curves $\delta
_{T}^*\left( C\right) $ are in excellent agreement with the experiments. The
fact that values of the fitting parameter ( $\chi \simeq .25$ for AFP--I,
and $\chi \approx .33$ for AFP--III) are physically reasonable and close to
each other gives an additional support to our mechanism.

Note that according to Eq. (\ref{DeltaT1}), the activity  is
strongly dependent on the smaller  dimension of the AFP. 
This may be  useful for the  design of
synthetic  AFP's. E. g., by using ring--shaped molecules
the $\delta_T^*$ could be increased
since the effective size is set by the largest dimension,  the 
radius of the ring, in this case.  Our results are  consistent with the
measured  $\delta _{T}^*\left(C\right) $    for Antifreeze Glycoproteins (AFGP).
However, their molecular architecture and conformations are more complicated
than those we have discussed,  and their analysis would go beyond the
scope of this work. 

Finally,  we can compare our results with the  
related problem of the motion of a pulled elastic interface in a 
medium with static obstacles to interface motion.  This is of interest  in the 
description  of the kinetics of domain walls, charge density waves and flux lines 
in superconductors \cite{Kardar98}. The interplay of long-range  elastic 
coupling and local pinning results in a
pinning-depinning transition  at a critical value of the pulling 
force, and near the transition point, the average speed of the 
interface goes continuously to zero.
Our model has pinning of 
the  interface  by the AFP molecules, and coupling due to surface 
tension,  but the transition is discontinuous. The 
difference is that the AFP's
are not stationary. Their arrival at the interface is controlled by 
the adsorption process and is independent of the advance of the 
interface. To emphasize this difference, we call our model 
`kinetic pinning'. There should be a crossover between the  
static  pinning  and kinetic pinning   if 
diffusion of the obstacles is taken into account.

We would like to thank Charles Knight  and E. Brener for useful conversations.
LMS  acknowledges partial support by 
NSF grant No.  DMS-0244419 and also
the hospitality of the Kavli Institute for Theoretical Physics 
where this research was supported in part by NSF grant No. PHY99-07949.

\bibliography{icebib}

\end{document}